\documentclass[12pt, fleqn]{article}
\usepackage{latexsym,amsfonts,amssymb}

\usepackage{amsfonts}
\usepackage{amssymb}
\usepackage{amsbsy}
\usepackage{amsmath}
\usepackage{epsf}
\usepackage{graphicx}

\newtheorem{theo}{Theorem}
\newcommand{\bt}{\begin{theo}}
\newcommand{\et}{\end{theo}}
\newcommand{\bd}{\begin{displaymath}}
\newcommand{\ed}{\end{displaymath}}

\newcommand{\be} {\begin{equation}}
\newcommand{\ee} {\end{equation}}
\newcommand{\ba} {\begin{array}}
\newcommand{\ea} {\end{array}}

\newcommand{\p} {\partial}

\begin{document}

\begin{center}
 {\Large \bf A mathematical model \\
for fluid-glucose-albumin  transport \\
 in peritoneal dialysis}

{\bf R. Cherniha~$^1,^2$,  J. Stachowska-Pietka~$^\dag$  and  J. Waniewski~$^\dag$ }

{\it $^1$~Institute of Mathematics,  NAS of Ukraine,\\
Tereshchenkivs'ka Street 3, 01601 Kyiv, Ukraine\\
$^2$~Department  of  Mathematics,
     National University
     `Kyiv-Mohyla Academy', \\ 2 Skovoroda Street,
     Kyiv  04070 ,  Ukraine}\\
 \texttt{cherniha@gmail.com}\\
{\it  $^\dag$~Institute of Biocybernetics and Biomedical
Engineering,
PAS,\\Ks. Trojdena 4, 02 796 Warszawa, Poland}\\
\texttt{jstachowska@ibib.waw.pl} \\
\texttt{jacekwan@ibib.waw.pl}

\end{center}

\begin{abstract}

A mathematical model for fluid and solute transport  in peritoneal dialysis is
constructed. The model is based on a three-component nonlinear
system of two-dimensional partial differential equations for fluid, glucose and albumin transport with the
relevant boundary and initial conditions.
Its aim is to model  ultrafiltration  of  water combined with inflow of glucose to the tissue and  removal  of
albumin
 from the body
during dialysis, and it does this by finding the spatial distributions of glucose and albumin
concentrations and hydrostatic pressure. The model  is developed in   one spatial dimension  approximation and
a governing
equation for each of the variables is derived from physical principles.
Under certain
assumptions the model  are  simplified
 with  the aim of obtaining
 exact formulae for
spatially non-uniform steady-state solutions.
 As the result, the exact
formulae for the  fluid fluxes from blood to tissue and  across the tissue are constructed together with two linear
autonomous ODEs for glucose and albumin concentrations in the tissue. The obtained analytical results are checked for their
applicability for the description of fluid-glucose-albumin  transport  during
peritoneal dialysis.

\end{abstract}



\newpage
\centerline{{\bf 1 Introduction}}

 Peritoneal dialysis is a life
saving treatment for chronic patients with end stage renal disease
(Gokal R and   Nolph 1994). The peritoneal cavity, an empty space that separates bowels, abdominal muscles and other organs in the abdominal cavity, is applied as a container for dialysis fluid, which is infused there through a permanent catheter and left in the cavity for a few hours. During this time  small metabolites (urea,
creatinine) and other uremic toxins diffuse from blood that perfuses the tissue layers close to the peritoneal cavity to the dialysis fluid,
and finally are removed together with
the drained fluid. The treatment cycle (infusion, dwell, drainage) is repeated several times every day. The
peritoneal transport occurs between dialysis fluid in the peritoneal
 cavity and blood passing down the capillaries in tissue surrounding the
peritoneal cavity. The capillaries are distributed within the tissue
at different distance from the tissue surface in contact with
dialysis fluid. The solutes, which are transported between blood and
dialysis fluid, have to cross two transport barriers: the capillary
wall and a tissue layer (Flessner 2006). Typically, many solutes are transported from
blood to dialysate, but some solutes such as for example an osmotic agent   (it is typically glucose), that is present in high
concentration in dialysis fluid, are transported in the opposite direction, i.e., to the  blood.
This kind
of transport system happens also in other medical treatments, as
local delivery of anticancer medications, and some experimental or
natural physiological phenomena (see below). Typically, to take into account spatial properties of these systems, a distributed approach is applied. The first applications of the distributed model were limited to the diffusive transport of gases between blood and artificial gas pockets within the body (Piiper et al 1962), between subcutaneous pockets and blood (Van Liew 1968, Collins 1981), and the transport of heat and solutes between blood and tissue (Perl 1963, Pearl 1962). The applications of the distributed approach for   modeling of the diffusive transport of small solutes include
 the description of the transport from cerebrospinal fluid to the brain (Patlak 1975), delivery of drugs to the human bladder during intravesical chemotherapy, and drug delivery from the skin surface to the dermis in normal and cancer tissue (Gupta et al 1995; Wientjes et al. 1993; Wientjes et al. 1991). Finally, the distributed approach was also proposed for the theoretical description of fluid and solute transport in solid tumors (Baxter and Jain 1989, 1990, 1991). The mathematical description of these
systems was obtained using partial differential equations based on
the simplification that capillaries are homogeneously distributed
within the tissue.
 Experimental evidence confirmed the good
applicability of such models (see, for example, the papers  (Waniewski et al. 1996a,1996b; Smit et al 2004a; Flessner 2006; Parikova et al. 2006; Waniewski et al. 2007; Stachowska-Pietka et al.  2012) and references therein).

An important objective of peritoneal dialysis is to remove excess water
from the patient
(Gokal R and   Nolph 1994).    The typical values of the water  ultrafiltration measured  during  peritoneal dialysis  are  $10 - 20 $ $mL/min$ (Heimb\"urger et al. 1992; Waniewski et al. 1996a,1996b; Smit et al 2004a; Smit et al 2004b). This is gained by inducing
osmotic pressure in dialysis fluid by adding a solute (called osmotic agent) in high
concentration. The most often used osmotic agent is glucose. This medical
application of high osmotic pressure is unique for peritoneal
dialysis. The flow of water from blood across the tissue to dialysis fluid in the peritoneal cavity carries solutes of different size, including large proteins, and adds a convective component to their diffusive transport.

Mathematical description of fluid and solute transport between blood and dialysis
fluid in the peritoneal cavity has not been formulated fully yet, in spite of the well
 known basic physical laws for such transport.
  The complexity of the peritoneal fluid transport modelling comes mainly from the fact that, whereas diffusive transport of small solutes is linear, process of water removal during peritoneal dialysis by osmosis is nonlinear.  A first formulation of the general distributed model for combined solute and fluid transport was proposed  by Flessner et al. (1984) and applied later for the description of the peritoneal transport of small molecules (Flessner et al. 1985).

  The next  attempt to model fluid and solute transport did not result in a satisfactory description.
  It was assumed in that model that mesothelium is a very efficient osmotic barrier for glucose with the same transport characteristics as endothelium (Seams et al, 1990). The assumption resulted in negative interstitial hydrostatic pressures during osmotically driven ultrafiltration from blood to the peritoneal cavity during peritoneal dialysis (Seams et al, 1990).  This prediction was shown to contradict the experimental evidence on positive interstitial hydrostatic pressure during ultrafiltration period of peritoneal dialysis (Flessner 1994). Moreover, the mesothelium being a very permeable layer cannot provide enough resistance to small solute transport to be an osmotic barrier for such solutes as glucose (Flessner 1994; Czyzewska et al. 2000; Flessner 2006).

  Recent mathematical, theoretical and numerical studies
   introduced new concepts on peritoneal transport and yielded
   better description of particular processes such as pure water transport, combined osmotic fluid flow and small solute transport, or water and proteins transport
   (Flessner 2001, Cherniha and  Waniewski 2005;
   Stachowska-Pietka et al. 2006,2007;    Cherniha et al. 2007; Waniewski et al.2007, Waniewski et al.2009).
   The recent study (Stachowska-Pietka et al.  2012) addresses again a combined transport of fluid (water)  and several small solutes.
     However, the problem of
     a combined description of osmotic ultrafiltration to the
     peritoneal cavity, absorption of osmotic agent from the
     peritoneal cavity and leak of macromolecules
     (e.g., albumin) from blood to the peritoneal cavity
     has not been addressed yet.
     Therefore, we present here
     an extended model for these phenomena and investigate
     its mathematical structure.
In particular, the present study is aimed on investigation of some basic questions concerning the role of various transport
components, as osmotic and oncotic gradients and hydrostatic pressure gradient.
 It should be stressed that the oncotic gradient leading to  leak of macromolecules from blood to the peritoneal cavity has opposite sign to the osmotic gradient, hence,  their combination may lead to new effects, which do not arise  in the case of the simplified models mentioned above.

The paper is organized as follows. In section 2, a mathematical
model of glucose and albumin transport in peritoneal dialysis is
constructed. In section 3, non-uniform   steady-state solutions of
the model are constructed and their properties are investigated.
Moreover, these solutions are tested for the real parameters that
represent clinical treatments of peritoneal dialysis. The results
are compared with those derived by numerical simulations for
simplified models
(Cherniha et al. 2007, Waniewski et al. 2007).
Finally, we present some
conclusions and discussion in the last section.



\medskip
\centerline{\textbf{2. Mathematical model}}

Here we present new model  of fluid and solute  transport in peritoneal dialysis. The model   is developed in   one spatial dimension  with $x = 0 $ (see the vertical orange line in Fig.1) representing the boundary of the peritoneal cavity
and $ x = L$  representing the end of the tissue  surrounding the
peritoneal cavity, see (Stachowska-Pietka et al. 2012) for the discussion of the assumptions involved in this approach.

 The mathematical description of
transport processes within the tissue consists in local balance of
fluid volume and solute mass. For incompressible fluid, the change
of volume may occur only due to elasticity of the tissue.
The fractional fluid
void volume, i.e. the volume occupied by the fluid in the
interstitium (the rest of the tissue being cells and macromolecules
forming the solid structure of the interstitium) expressed per one unit volume of the whole
tissue, is denoted by $\nu(t,x)$, and its time evolution is described as:
 \be \label{1.1} {{\partial \nu} \over
{\partial t}} = - {{\partial j_U } \over {\partial x}} + q_U - q_l
\ee where $j_U(t,x)$ is the volumetric fluid flux across the tissue
(ultrafiltration), $q_U(t,x)$ is the density of volumetric fluid
flux from blood capillaries to the tissue, and  $q_l$ is the density of volumetric fluid
flux from  the tissue to the lymphatic vessels (hereafter we
assume that it is a known positive constant, nevertheless it can be
also a function of hydrostatic pressure ( Stachowska et al, 2006, 2012)).
 Similarly to  many  distributed models, our model involves the spreading of the source within the whole tissue as an approximation to the discrete structure of blood and lymphatic capillaries.

The independent variables are time $t$ and the distance $x$ within the tissue from
the tissue surface in contact with dialysis fluid in the direction perpendicular to this surface (flat geometry of
the tissue is here assumed with finite width, see below). The solutes, glucose and albumin, are
distributed only within the interstitial fluid (or part of it, see below), and their
concentrations in this fluid are denoted by $C_G(t,x)$ and
$C_A(t,x)$, respectively. The equation that describes the local
changes of glucose  amount in the tissue, $\nu C_G$, is:
 \be
\label{1.2} {{\partial (\nu C_G )} \over {\partial t}} =  -
{{\partial j_G } \over {\partial x}} + q_G, \ee
 where $j_G(t,x)$ is
glucose flux through the tissue, and $q_G(t,x)$ is the density of
glucose flux from blood.
 The cellular uptake of absorbed glucose is not taken into account in  equation  (\ref{1.2}) because this process  leads to   a small correction to the bulk absorption of glucose to the capillaries. So, we neglect the intracellular changes that were noted experimentally (Zakaria et al. 2000).

Similarly, the equation that describes the local changes of albumin
amount in the tissue, $\alpha \nu C_A$, is:
 \be
\label{1.2-a} {{\partial (\alpha \nu C_A )} \over {\partial t}} =  -
{{\partial j_A } \over {\partial x}} + q_A, \ee
 where $j_A(t,x)$ is
albumin  flux through the tissue, $q_A(t,x)$ is the density of
albumin flux from blood. The coefficient $\alpha<1$ takes into
account  that only a part of the  fractional fluid void volume $\nu$ that is available for fluid, is
accessible for albumin because of its  large molecular size
  (Flessner  2001; Stachowska-Pietka et al. 2007).
  In other words, the inclusion of the term $\alpha \nu$ in  (\ref{1.2-a}) implies that  $C_A(t,x)$  is
the concentration of albumin in that part of the interstitium
 across which the
albumin molecules can pass.
 In the general case,  equation  (\ref{1.2-a}) involves a new fluid void volume function $ \nu_A(t,x)$,  which depends  on the hydrostatic pressure similarly to  the function $\nu$ (see below) and satisfies the inequality $ \nu_A< \nu$. Hereafter we set $ \nu_A=\alpha \nu$ for simplicity.

 The flows of fluid and solutes through the tissue are
described according to linear non-equilibrium thermodynamics.
Osmotic pressure of glucose and oncotic pressure of  albumin  are
described by van't  Hoff  law, i.e. assuming that corresponding pressures are proportional to the relevant
concentrations.

The
fluid  flux across the tissue is generated by hydrostatic,
osmotic and oncotic (i.e., osmotic pressure of large proteins) pressure gradients:
 \be \label{1.3} j_U = - \nu K{{\partial P} \over
{\partial x}} + \sigma _{TG} \nu KRT{{\partial C_G } \over {\partial
x}}+ \sigma _{TA}
 \nu KRT{{\partial C_A } \over {\partial x}},
\ee
where $K$ is the hydraulic conductivity
of tissue that is assumed constant  for simplicity( $K$ may also depend on the  pressure $P$), $R$ is the gas constant, $T$ is absolute temperature,
and $\sigma_{TG}$ and $\sigma_{TA}$ are the Staverman reflection coefficients for glucose and albumin in tissue, respectively.  The Staverman reflection coefficient $\sigma $ is a thermodynamic parameter and describes the effectiveness of osmotic pressure in selectively permeable membrane: if $\sigma = 0$ then no osmotic pressure can be induced by this solute across the membrane, and if $\sigma = 1$ the maximal theoretically possible osmotic effect can be induced (ideal semi-permeable membrane).  The intermediate values of $\sigma$ represent  non-ideal  semipermeable membranes.  The book (Currant, Katchalsky 1965) well addresses the problem of the Staverman reflection coefficients.

  The density of fluid flux from blood to tissue is generated, according to Starling law, by the hydrostatic, osmotic and oncotic
pressure differences between blood and tissue:
 \be \label{1.4} q_U = L_p a(P_{B} - P)- L_p a\sigma_{G} RT(C_{GB} -
 C_G)-
  L_p a\sigma_{A} RT(C_{AB} - C_A), \ee
where $P(t,x)$ is hydrostatic  pressure, $L_p a$ is the hydraulic conductance
 of the capillary wall,
   $P_B$ is the hydrostatic pressure of blood,   $C_{GB} $ and $C_{AB} $  are  glucose and albumin concentrations
in blood, and  $\sigma_{G}$ and $\sigma_{A}$ are the Staverman reflection coefficients for glucose and albumin in the capillary wall, respectively.  In contrary to other parameters, there is an unsolved  problem of the values of $\sigma_{G}$ and $\sigma_{A}$.  In particular, the values of $\sigma_{G}$  were found low (about $0.005 - 0.03$) in many experiments  in contrast to some newer experimental data that suggest the values close to $0.5$ (see the discussion of this controversy in (Waniewski et al. 2009, Stachowska-Pietka et al.  2012)).
We also assume that blood concentrations of glucose and albumin   are constant
according to the clinical and experimental data that demonstrate only negligible variation of these concentrations during peritoneal dwell of dialysis fluid (Heimburger et al. 1992).  This observation is related to a quasi-continuous mode of continuous ambulatory peritoneal dialysis with fluid exchanges every few hours and was applied in most previous theoretical and numerical studies on peritoneal dialysis.

  The glucose
   flux across the tissue
is composed of diffusive component (proportional to glucose
concentration gradient) and convective component (proportional to
glucose concentration and fluid flux):
\be \label{1.5} j_G  = -
\nu D_G {{\partial C_G } \over {\partial x}} + S_{TG} C_G j_U. \ee
where $D_G$ is the diffusivity  of glucose in tissue,
$S_{TG}$ is the sieving coefficients of glucose in tissue. According to non-equilibrium thermodynamics, $S_{TG} = 1-\sigma_{TG}$ for homogenous membrane ( Katchalsky, Currant 1965).

  The density of glucose flux between blood and the tissue describes the number of moles of glucose per unit total
volume of tissue per unit time that move between  blood and   tissue.
  It
   has diffusive
component (proportional to the difference of glucose concentration
in blood, $C_{GB}$, and glucose concentration in tissue, $C_G$),
convective component (proportional to the density of fluid flow
from blood to tissue, $q_U$) and the component that represents  lymphatic
absorption of solutes (proportional to the density of volumetric lymph flux, $q_l$):
 \be \label{1.6} q_G = p_G a(C_{GB}-C_G )+S_{G}q_U C_G
  - q_lC_G. \ee
where $p_G a$ is the diffusive permeability of the capillary wall for glucose.

  In a similar way,  the albumin
 flux across the tissue,  $j_A(t,x)$, and the density of albumin
flux  to tissue, $q_A(t,x)$, can be described as:
\be \label{1.5-a} j_A  = -
\alpha \nu D_A {{\partial C_A } \over {\partial x}} + S_{TA} C_A
j_U, \ee \be
 \label{1.6-a} q_A = p_A a(C_{AB}-C_A )+S_{A}q_U C_A
  - q_lC_A. \ee
where $S_{TA}= 1-\sigma_{TA}$ is the sieving coefficient of albumin in
tissue, $S_{A} = 1-\sigma_{A}$ is the sieving coefficient of glucose and
albumin in the capillary wall, $D_A$ is the diffusivity of albumin  in tissue, and $p_A a$ is the diffusive permeability of the capillary wall for
albumin.

  The  typical  values of  the model parameters are listed in Table 1.

Equations (\ref{1.1})-(\ref{1.2-a}) together with equations
(\ref{1.3})-(\ref{1.6-a}) for flows form a system of three nonlinear
partial differential equations with four variables: $\nu, P, C_A$,
and $C_G$. Therefore, an additional, constitutive, equation is
necessary, and this is the equation describing how fractional fluid
void volume, $\nu$, depends on interstitial pressure, $P$. This
dependence can be established  using data from experimental studies
  (Stachowska-Pietka et al. 2006). It turns out that \be \label{1.7-a} \nu =
F(P), \ee where $F$ is a monotonically non-decreasing bounded
function with the limits: $F \to \nu_{\min}$ if $P \to P_{\min}$ and
$F \to \nu_{\max}$ if $P \to P_{\max}$ (particularly, one may take
$P_{\min}=-\infty, \ P_{\max}=\infty$).
Here $\nu_{\min}<1 $ and $ \nu_{max}<1 $ are empirically measured
constants.
 For example, the following  analytical form for the
function $F$   based on experimental data taken from (Zakaria et al. 1999)
\[
 \nu(P) = \nu_{\min}+ \frac
{\nu_{\max}-\nu_{\min}}{1+\left( \frac{\nu_{\max }- \nu_{\min }}
{\nu_0 -\nu_{\min}} -1 \right) e^{ - bP} }, \quad b>0 \]
was used in
  (Stachowska-Pietka et al. 2006; Cherniha et al. 2007).

Boundary conditions for a tissue layer of width $L$ impermeable at
$x = L$ and in contact with dialysis fluid at $x = 0$ are:
 \be \label{1.8}  x = 0: \ P
= P_D, \quad  C_G = C_{GD}, \quad  C_A = C_{AD} \ee
 \be \label{1.8*}  x = L: \
{{\partial P} \over {\partial x}} = 0,
 \quad {{\partial C_G} \over {\partial x}} = 0, \quad {{\partial C_A} \over {\partial x}} = 0.\ee
 Generally speaking, intraperitoneal pressure  $P_D$, glucose $ C_{GD}$  and albumin $ C_{AD}$  concentrations in the peritoneal cavity  may depend on time. However, experimental data and theoretical studies suggest that they change at low rate compared to the rate of transport processes in the tissue (Stachowska-Pietka et al. 2006,2007;Waniewski 2007). Therefore, we may assume that $P_D$, $ C_{GD}$ and $ C_{AD}$ are constant for some time period and assess the steady-state solution for these particular boundary conditions that may be considered as approximated quasi steady-state solution for the full model of peritoneal dialysis with time-dependent  boundary conditions. This approximation was applied previously    for  the model with variable boundary conditions for small solutes (as glucose) and water transport (but without proteins, as albumin), see (Dedrick  1981; Flessner et al.1984, Flessner et al. 1985; Seames et al. 1990;
    Stachowska-Pietka et al.  2006, Stachowska-Pietka et al.  2007; Waniewski 2001, 2002;
  Waniewski et al.   2009).

 Initial conditions describe equilibrium within the tissue
without any contact with dialysis fluid: \be \label{1.9}
t = 0: \ P = P^*, \ C_G = C^*_G, \ C_A = C^*_{A},
\ee
where  $P^*, \  C^*_G, $ and $  C^*_{A}$ are some non-negative  values, which will be estimated  below.

Note  that equations (\ref{1.1})-(\ref{1.7-a}) can be
united into three nonlinear partial differential equations (PDEs)
for hydrostatic pressure $P(t,x)$, glucose
concentration $C_G(t,x)$ and albumin concentration $C_A(t,x)$. Thus, these three
PDEs together with boundary and initial conditions
(\ref{1.8})-(\ref{1.9}) form a nonlinear  boundary-value problem.
  Possible values of the parameters arising in this problem
  are presented in Table 1 (see the relevant comments in Section 4).

The fluid flux $j_U(t,x)$
 at $ x = 0$  describes the net ultrafiltration flow, i.e., the exchange of fluid 
   between the tissue and the peritoneal cavity across the peritoneal surface and therefore directly the efficiency of removal of water
    during peritoneal dialysis. The assessment of ultrafiltration flow is important from practical point of view because the low values of this flow  in some patients indicate that some problems with osmotic fluid removal occur, which   may finally result in the failure of the therapy  (Parikova 2006).

\medskip

\centerline{\textbf{3. Steady-state solutions of the  model and
their applications}}

First of all, we consider the special case, with  tissue  in its physiological state without dialysis, and, therefore, no transport to the peritoneal cavity occurs. In this case  the boundary
conditions at $x=0$ given by Eq. (\ref{1.8})  are replaced by zero Neumann conditions, and
 the steady-state solution can be easily  found because  it does not depend on $x$. In fact solving algebraic equations \be
\label{2.0} q_U - q_l =0, \quad q_G =0, \quad q_A =0, \ee one easily
obtains the spatially uniform steady-state solution \be \label{2.*}
\ba{l}\medskip C_G^*= \frac{p_G a}{p_G
a+\sigma_{G}q_l} C_{GB}\\ \medskip C_A^*= \frac{p_Aa}{p_A a+\sigma_{A}q_l}
C_{AB}\\
P^* = P_B -q_l\Bigl(\frac{1}{L_pa}+RT \Bigl(\frac{\sigma_G^2 C_{GB}}{p_G
a+\sigma_{G}q_l}+\frac{
\sigma_A^2 C_{AB}}{p_A
a+\sigma_{A}q_l}\Bigr)\Bigr).\ea \ee In the case $q_l=0$, i.e.,
zero flux from the tissue to the lymphatic vessels, formulae
(\ref{2.*}) produce \be \label{2.**} C_G^*= C_{GB}, \quad  C_A^*=
C_{AB}, \quad P^* = P_B,  \ee otherwise \be \label{2.***} C_G^*<
C_{GB}, \quad  C_A^* < C_{AB}, \quad P^* < P_B.  \ee
This uniform solution describes the system in equilibrium if no dialysis is performed, and therefore we may use the values $ P^*, \  C^*_G, $ and $  C^*_{A}$
calculated above as the initial profile for simulation of the transport processes after the initiation of dialysis (see formulae  (\ref{1.9}) ).

To find spatially non-uniform  steady-state solutions,  we reduce Eqs.
(\ref{1.1})-(\ref{1.2-a}) to an equivalent form by introducing scaled
non-dimensional independent and dependent variables (except for $\nu$ that is a non-dimensional variable) \be \label{2.1}
x^*=\frac xL, \quad t^*=\frac{KP_Dt}{L^2}, \ee
 \be \label{2.2}
p(t^*,x^*)= \frac{P}{P_D},
 \quad u(t^*,x^*) =
\frac{C_G-C_{GB}}{C_{GD}-G_{GB}}, \quad
w(t^*,x^*)=\frac{C_A}{C_{GD}-G_{GB}}.
 \ee
Thus, after rather simple calculations and taking into account
 Eqs. (\ref{1.3}),  (\ref{1.5}), and (\ref{1.5-a}), one obtains
 Eqs. (\ref{1.1})-(\ref{1.2-a}) in  the
  form (hereafter
 upper index $*$
is omitted) \be \label{2.3}
\frac{\p\nu}{\p t}=
\frac{\p}{\p x} \Bigl({\nu{\partial p} \over {\partial
x}}\Bigr) - t_0\sigma _{1} \frac{\p}{\p x} \Bigl({\nu{\partial u} \over
{\partial x}}\Bigr)- t_0\sigma _{2} \frac{\p}{\p x} \Bigl({\nu{\partial
w} \over {\partial x}}\Bigr)  +t_0( q_U - q_l),\ee
\medskip
 \[
 \frac{\p(\nu u)}{\p t} + \sigma_{TG}u_0 \frac{\p\nu}{\p t}=
d _{1}t_0 \frac{\p}{\p x} \Bigl({\nu{\partial u} \over {\partial
x}}\Bigr)+S_{TG}\frac{\p}{\p x} \Bigl({u\nu{\partial p}
\over {\partial x}}\Bigr) - S_{TG}t_0\sigma _{1} \frac{\p}{\p x}
\Bigl({u\nu{\partial u} \over {\partial x}}\Bigr)\] \be \label{2.4}
- S_{TG}t_0\sigma _{2} \frac{\p}{\p x} \Bigl({u\nu{\partial w} \over
{\partial x}}\Bigr)+t_0(S_Gu+u_0(S_G-S_{TG}))q_U
-t_0b_1u-t_0\sigma_{TG}u_0q_l,\ee\medskip
 \[
 \frac{\p(\alpha \nu w)}{\p t} -S_{TA}w_0\frac{\p\nu}{\p t}=
d _{2}t_0 \frac{\p}{\p x} \Bigl({\nu{\partial w} \over {\partial
x}}\Bigr)+S_{TA}\frac{\p}{\p x} \Bigl({w^*\nu{\partial
p} \over {\partial x}}\Bigr) - S_{TA}t_0\sigma _{1} \frac{\p}{\p x}
\Bigl({w^*\nu{\partial u} \over {\partial x}}\Bigr)\] \be
\label{2.5} - S_{TA}t_0\sigma _{2} \frac{\p}{\p x}
\Bigl({w^*\nu{\partial w} \over {\partial
x}}\Bigr)+t_0(S_Aw-S_{TA}w_0))q_U
-t_0b_2w^*-t_0\sigma_{TA}w_0q_l,\ee
 where \be \label{2.6} \ba{l}
 \medskip
q_U=  \beta \Bigl(
\frac{1}{t_0}(p_0-p)+\frac{\sigma_{G}\sigma_{1}}{\sigma_{TG}}u+
\frac{\sigma_{A}\sigma_{2}}{\sigma_{TA}}w^*
\Bigr),  \quad   \beta=\frac{L_paL^2}{K},\\
  \sigma_1=\sigma_{TG}KRT \frac{C_{GD}-G_{GB}}{L^2}, \quad
  \sigma_2=
  \sigma_{TA}KRT  \frac{C_{GD}-G_{GB}}{L^2},\\
d_1=\frac{D_{G}}{L^2}, \quad d_2=\frac{\alpha D_{A}}{L^2}, \\
 b_1= p_Ga+q_l, \quad  b_2= p_Aa+q_l, \\
 u_0=\frac{C_{GB}}{C_{GD}-G_{GB}},   \quad
 w_0=\frac{C_{AB}}{C_{GD}-G_{GB}}, \quad p_0 = \frac{P_B}{P_D}\\

t_0 =
 \frac{L^2}{KP_D},\quad w^*=w-w_0, \quad
 \ea \ee

We want to find  the steady-state solutions of Eqs.
(\ref{2.3})-(\ref{2.5}) satisfying the boundary conditions
(\ref{1.8})-(\ref{1.8*}). They take the form \be \label{2.7} x = 0:
\ p = 1, \quad u = 1, \quad  w = \frac{C_{AD}}{C_{GD}-G_{GB}} \ee
 \be \label{2.8}  x = 1: \
{{\partial p} \over {\partial x}} = 0,
 \quad {{\partial u} \over {\partial x}} = 0, \quad {{\partial w} \over {\partial x}} = 0.
\ee for the non-dimensional variables.

\medskip

Note  that to find the steady-state solutions  Eqs. (\ref{2.3})-(\ref{2.5}) can be  reduced to the  system of
ordinary differential equations (ODEs)
\be \label{2.10}  \frac{1}{t_0}\frac{d}{d x} \Bigl({\nu{d
p} \over {d x}}\Bigr) - \sigma _{1} \frac{d}{d x} \Bigl({\nu{d u}
\over {d x}}\Bigr)- \sigma _{2} \frac{d}{d x} \Bigl({\nu{d w} \over
{d x}}\Bigr) + q_U - q_l = 0,\ee
\medskip
 \[
d _{1} \frac{d}{d x} \Bigl({\nu{d u} \over {d
x}}\Bigr)+\frac{S_{TG}}{t_0}\frac{d}{d x} \Bigl({u\nu{d p} \over {d
x}}\Bigr) - S_{TG}\sigma _{1} \frac{d}{d x} \Bigl({u\nu{d u} \over
{d x}}\Bigr)\] \be \label{2.11} - S_{TG}\sigma _{2} \frac{d}{d x}
\Bigl({u\nu{d w} \over {d x}}\Bigr)+(S_Gu+u_0(S_G-S_{TG}))q_U
-b_1u-\sigma_{TG}u_0q_l =0 ,\ee\medskip
 \[
d _{2} \frac{d}{d x} \Bigl({\nu{d w} \over {d
x}}\Bigr)+\frac{S_{TA}}{t_0}\frac{d}{d x} \Bigl({w^*\nu{d p} \over
{d x}}\Bigr) - S_{TA}\sigma _{1} \frac{d}{d x} \Bigl({w^*\nu{d u}
\over {d x}}\Bigr)\] \be \label{2.12} - S_{TA}\sigma _{2} \frac{d}{d
x} \Bigl({w^*\nu{d w} \over {d
x}}\Bigr)+(S_Aw-S_{TA}w_0)q_U -b_2w^*-\sigma_{TA}w_0q_l =
0.\ee

Non-linear system of ODEs (\ref{2.10})-(\ref{2.12}) is still
very complex  and cannot be  integrated in the case of   arbitrary coefficients. Thus,
we look for the correctly-specified coefficients, for which this system
can be simplified. It can be noted that the relations \be
\label{2.13} S_A=S_{TA}, \quad S_G=S_{TG} \ee lead to an essential
(it means that automatically $\sigma_{G}= \sigma_{TG}, \ \sigma_{A}= \sigma_{TA}$) simplification of this system.
 This assumption is introduced for mathematical reason only: a specific symmetry of the equations allows for much easier  rigorous  analysis. On the other hand, it is shown in the next section that even in this special case the solutions of the model are qualitatively/quatitatively  similar to those obtained via  other simplified models, which do not use this assumption.

So, using assumption (\ref{2.13}),
expressions for $q_U$ from (\ref{2.6}) and   $j_U$ from (\ref{1.3}),
rewritten in non-dimensional variables
 \be \label{2.13a} j_U =  L\nu \Bigl(-\frac{1}{t_0}{{\partial p} \over
{\partial x}} + \sigma _{1}  {{\partial u } \over {\partial x}}+
\sigma _{2} {{\partial w } \over {\partial x}}\Bigr), \ee
 one obtains the relation
\be \label{2.15} j_U = \frac{K\nu}{L_paL}\frac{dq_U}{d x} \ee
allowing to find $j_U$ provided
the  function $q_U$  is known. Using the formulae  (\ref{2.13}) --
 (\ref{2.15}),
the nonlinear  ODE system  (\ref{2.10}) -- (\ref{2.12}) can be simplified to the form
 \be \label{2.14}
\frac{1}{\beta}\frac{d}{d x}\Bigl(\nu{{d q_U} \over {d x}}\Bigr)- q_U + q_l=0 \ee
\medskip
 \be \label{2.14*}d _{1} \frac{d}{d x} \Bigl(\nu{{d u} \over {d
x}}\Bigr)-\frac{S_G}{\beta} \frac{d}{d x}\Bigl(\nu u{{d q_U} \over {d x}}\Bigr) +S_Guq_U
-b_1u  - \sigma_{G}u_0q_l =0 ,\ee\medskip
\be \label{2.14**} d _{2} \frac{d}{d x} \Bigl(\nu{{d w} \over {d
x}}\Bigr)
-\frac{S_A}{\beta} \frac{d}{d x}\Bigl(\nu w{{d q_U} \over {d x}}\Bigr)+S_Awq_U
-b_2(w-w_0) - w_0q_l =0. \ee

The linear  semi-coupled system of ODEs  (\ref{2.15}) -- (\ref{2.14})  can be extracted to find the functions
 $q_U$ and $j_U$ provided the
function $\nu$ is known.
  However $\nu$ depends on pressure,
which is  also unknown function, and therefore we need to assume additional
restrictions on the function $F$ from  formula (\ref{1.7-a}).

\vspace{5mm}

Let us consider first case, in which we assume that
 $F$ is a constant function. This assumption was applied in many studies, especially for the description of small solute transport (Dedrick 1981, Flessner 1984, Waniewski 2001, 2002).
In this case  \be \label{2.16} \nu(x) = \nu_m, \ee where
$\nu_m$ is a positive constant.
Substituting (\ref{2.16}) into
system  (\ref{2.15})--(\ref{2.14}),  its general
solution can be found: \be \label{2.17} q_U = C_1e^{-\lambda x} + C_2e^{\lambda
x} + q_l, \ee \be \label{2.18} j_U = \frac L\lambda (-C_1e^{-\lambda
x} + C_2e^{\lambda x}),  \quad  \lambda=\sqrt{\frac{\beta}{\nu_m}}=
\sqrt{\frac{L_paL^2}{K\nu_m}}.\ee The arbitrary constants $C_1$ and
$C_2$  can be specified using the boundary conditions
(\ref{2.7}) -- (\ref{2.8}) since the functions $q_U$ and $j_U$ are
expressed via $p, \ u, \ w$ and its first-order derivatives (see
formulae (\ref{2.6}) and (\ref{2.13a})). Making rather simple
calculations, one obtains \be \label{2.19} C_1 = (q_0
-q_l)\frac{e^{2\lambda}}{1+ e^{2\lambda}}, \quad C_2 = (q_0
-q_l)\frac{1}{1+ e^{2\lambda}},\ee where \be \label{2.20} q_0 =
\beta \Bigl(
\frac{1}{t_0}(p_0-1)+\sigma_1+\sigma_{2}\frac{C_{AD}-C_{AB}}{C_{GD}-G_{GB}}\Bigr).\ee

Having the explicit  formulae for $q_U$ and $j_U$,
equations (\ref{2.14*}) and (\ref{2.14**})  can be
reduced to two linear autonomous ODEs: \be \label{2.21} d _{1}\nu_m
\frac{d^2u}{d x^2} +\frac{S_{G}}{\lambda}(C_1e^{-\lambda x} -
C_2e^{\lambda x})\frac{du}{d x}
- \kappa_1 u-u_{01} =0 \ee and \be \label{2.22}
d _{2}\nu_m \frac{d^2w}{d x^2} +\frac{S_{A}}{\lambda}(C_1e^{-\lambda
x} - C_2e^{\lambda x})\frac{dw}{d x}
-\kappa_2(w-w_0)-w_{01} =0 \ee
 with unknown
  functions
$u(x)$ and $w(x)$. Hereafter the notations  \be \label{2.23}
\kappa_1=p_Ga+\sigma_Gq_l, \quad
u_{01}=\sigma_{G}u_0q_l,
 \quad
\kappa_2=p_Aa+\sigma_Aq_l, \quad w_{01}=\sigma_{A}w_0q_l  \ee
are used. Note, the similarities in the structure of equations (\ref{2.21}) and (\ref{2.22}) However, to the   best of our knowledge, the  general solutions of ODEs
(\ref{2.21}) and (\ref{2.22})
are unknown . On the other hand,
since the unknown functions $u(x)$ and $w(x)$ should satisfy the
boundary conditions (\ref{2.7})--(\ref{2.8}), the corresponding
linear problems can be numerically solved using, for example, Maple
program package. Finally, using two expressions for $q_U$ from
(\ref{2.6}) and (\ref{2.17}), we obtain   the function   \be
\label{2.24}p(x)=
 p_0+t_0\sigma_{1}u+t_0\sigma_{2}(w-w_0)-
\frac{t_0}{\beta} \Bigl( C_1e^{-\lambda x} + C_2e^{\lambda x} +
q_l\Bigr). \ee

In the next section, the numerical non-uniform steady-state solutions based on the realistic  values of parameters arising in
the formulae derived above will be presented for this case i.e. with restrictions
(\ref{2.13}) and (\ref{2.16}).

\vspace{5mm}

Let us now consider the second type of restrictions  on function $\nu$. Instead of the rather restrictive assumption  (\ref{2.16}),  we
  examine the case when
the function $\nu$ is non-constant and satisfies the general conditions described after formula (\ref{1.7-a}).
According to the experimental data  the hydrostatic pressure during peritoneal dialysis is a decreasing function with respect to the distance $x$  from the peritoneal cavity (Flessner, 1994; Zakaria et al 1999, Zakaria et al 2000). Hence, function $F(p(x))$ is decreasing  (with respect to $x$ !) provided $p(x)$ is a spatially   non-uniform
 steady-state solution.
 The simplest case of such a pattern occurs when
$\nu$ is linear, monotonically decreasing function of $x$: \be \label{2.25}
\nu(p(x))\equiv \nu(x)=\nu_{max} -(\nu_{max}-\nu_{min})x, \quad x
\in [0,1]. \ee Substituting (\ref{2.25}) into (\ref{2.14}), we
obtain the linear ODE \be \label{2.26} (\nu_{max}
-(\nu_{max}-\nu_{min})x){{d^2 q_U} \over {d x^2}}  -
(\nu_{max}-\nu_{min}){{d q_U} \over {d x}}-
\beta(q_U - q_l) =0.\ee It can be shown
that by the substitutions:
\be \label{2.28} y^2 = 4\delta_*(\nu_* - x), \ q_V =
q_U - q_l, \ \nu_*=\frac{\nu_{max}}{\nu_{max}-\nu_{min}}>1, \
\delta_*= \frac
{\beta}{(\nu_{max}-\nu_{min})}>0\ee
the linear ODE (Eq. (\ref{2.26})) reduces to the  modified
Bessel equation of the zero order, see, e.g.,
 ( Polyanin A D and Zaitsev V F 2003) \be \label{2.27}
y^2\frac{d^2q_V}{dy^2} + y\frac{dq_V}{dy} -y^2q_V = 0. \ee
 The general
solution of Eq. (\ref{2.27}) is well-known.  Hence, using formulae
(\ref{2.28}), one obtains the  solution of Eq. (\ref{2.26}): \be
\label{2.29} q_U = C_1 I_0(2\sqrt{\delta_*(\nu_*-x)}) +
C_2K_0(2\sqrt{\delta_*(\nu_*-x)}) + q_l,\ee where $I_0$ and $K_0$
are the modified Bessel functions of the first and third kind,
respectively.

Substituting the obtained function $q_U$  into Eq. (\ref{2.15}) and using
the well-known relations between the Bessel functions
(Bateman 1974), we find the function: \be \label{2.30} j_U
=-L\sqrt{\frac{\nu_*-x}{\delta_*}}\Bigl( C_1
I_1(2\sqrt{\delta_*(\nu_*-x)}) -
C_2K_1(2\sqrt{\delta_*(\nu_*-x)})\Bigr),\ee where $I_1$ and $K_1$
are the modified Bessel functions of the first order.
Note  that, similarly  to previous case,
 the constants $C_1$ and $C_2$
 can be calculated  from the boundary conditions.    Omitting  rather
simple calculations, we present only the result:
\be \label{2.31}
C_1 = \frac{(q_0-q_l)K_1(2\sqrt{\delta_*(\nu_*-1)})}
{I_0(2\sqrt{\delta_*\nu_*})K_1(2\sqrt{\delta_*(\nu_*-1)})+
K_0(2\sqrt{\delta_*\nu_*})I_1(2\sqrt{\delta_*(\nu_*-1)})}, \ee \be
\label{2.32} C_2 = \frac{(q_0-q_l)I_1(2\sqrt{\delta_*(\nu_*-1)})}
{I_0(2\sqrt{\delta_*\nu_*})K_1(2\sqrt{\delta_*(\nu_*-1)})+
K_0(2\sqrt{\delta_*\nu_*})I_1(2\sqrt{\delta_*(\nu_*-1)})}, \ee where
$q_0$ is defined by (\ref{2.20}).

Thus, we have found  the explicit  formulae  for  $q_U$ and $j_U$.
Having
formulae (\ref{2.29}) -- (\ref{2.30}),   system of ODEs (\ref{2.14*})
-- (\ref{2.14**})   can be reduced to
two linear autonomous ODEs with the unknown  functions $u(x)$ and $w(x)$.
These equations possess the forms:
\be \label{2.33}\ba{l} d_{1}(\nu_{max}-\nu_{min})
\Bigl((\nu_*-x)\frac{d^2u}{d x^2}-\frac{du}{d x} \Bigr) \\
+\frac{S_{G}}{\sqrt\delta_*}\frac{d}{d x}\Bigl(\sqrt{\nu_*-x}(C_1
I_1(2\sqrt{\delta_*(\nu_*-x)}) -
C_2K_1(2\sqrt{\delta_*(\nu_*-x)}))u\Bigr)\\ +\Bigl( S_G(C_1
I_0(2\sqrt{\delta_*(\nu_*-x)}) +
C_2K_0(2\sqrt{\delta_*(\nu_*-x)}))-\kappa_1\Bigr)u-u_{01} =0 \ea\ee

and

\be \label{2.34} \ba{l} d_{2}(\nu_{max}-\nu_{min})
\Bigl((\nu_*-x)\frac{d^2w}{d x^2}-\frac{dw}{d x} \Bigr) \\
+\frac{S_{A}}{\sqrt\delta_*}\frac{d}{d x}\Bigl(\sqrt{\nu_*-x}(C_1
I_1(2\sqrt{\delta_*(\nu_*-x)}) -
C_2K_1(2\sqrt{\delta_*(\nu_*-x)}))(w-w_0)\Bigr) \ea \ee \[+\Bigl(
S_A(C_1 I_0(2\sqrt{\delta_*(\nu_*-x)}) +
C_2K_0(2\sqrt{\delta_*(\nu_*-x)}))-\kappa_2\Bigr)(w-w_0)-w_{01} =0\]

Although both equations are  linear second order ODEs with the
same structure, we could not find
their
general solutions because of their awkwardness. Thus, we
solve them numerically together with the boundary conditions
(\ref{2.7})--(\ref{2.8}) using the Maple program package. In the next
section,  realistic values of the parameters for formulae
(\ref{2.29})--(\ref{2.34}) will be selected  and applied in numerical simulations  to
calculate the non-uniform steady-state solutions.

\medskip
\textbf{Remark.} The results (with some misprints) of this section and Section 2 were briefly reported in (Cherniha  and Waniewski  2011).
\medskip

\centerline{\textbf{4. Numerical  results   and their biomedical interpretation }}

Here we present numerical results based on the formulae derived
in Section 3. Our aim is
to check
 whether they are
applicable for describing the fluid-glucose-albumin  transport  in
peritoneal dialysis.  The  parameters in these formulae were derived from experimental and clinical data and
applied in previous mathematical studies
(Van Liew 1968; Imholz et al. 1998;  Zakaria et al. 1999;  Flessner  2001; Smit et al 2004a; Smit et al 2004b; Waniewski 2001; Stachowska-Pietka et al. 2006; Cherniha et al. 2007, Stachowska-Pietka et al 2007, Waniewski et al. 2009 ).
Most of the parameters, especially those for water and glucose were derived from experimental data or obtained by fitting the distributed model to clinical data and are discussed in detail in a recent paper (Stachowska-Pietka et al. 2012).  Some phenomena, as vasodilatation and change of the parameters of interstitial transport with the change of tissue hydration, were not included into our modeling because our objective was the mathematical analysis of the model, so its structure should be simplified. Nevertheless, the model covers all basic transport phenomena and provides a good background for further modifications. However, only numerical studies are available for such extended models, see (Smit et al 2004a, Smit et al 2004b, Stachowska-Pietka et al. 2006). Furthermore, some parameters without firmly established experimental values (Staverman reflection coefficients, were varied to check their impact on the results of modelling . The diffusivity of albumin in the interstitium is  not well known, but it is much lower than interstitial diffusivity of glucose, see (Waniewski 2001, Stachowska-Pietka et al. 2012) for more details.

The values of parameters and absolute
constants applied in numerical simulations are listed in Table 1.


Let us consider the first  case  of  constant fractional fluid void volume, i.e.,
 with  restrictions (\ref{2.13}) and
(\ref{2.16}).    We remind the reader that the  assumption $F$ is a constant was applied in many studies and this implies that $\nu(x)$ is also a constant.  It seems to be reasonable to
set $\nu_m= (\nu_{max}+\nu_{min})/2 = 0.26
 $, i.e., we assume that
the fractional fluid void volume at the  steady-state stage of the
peritoneal transport is an intermediate value between its maximum and
minimum.
In order to compare  the  numerical results obtained here with those for osmotic peritoneal transport obtained earlier, in which albumin transport was not considered, we neglect the oncotic pressure as a driving fluid force across the tissue, i.e., we put the Staverman reflection coefficients for albumin  $\sigma_{TA}=\sigma_{A}=0$.
It means that  the fluid flux across  tissue,  $j_U$,  and the fluid flux
from blood to tissue, $q_U$, (see formulae  (\ref{1.3}) and
(\ref{1.4})) do not depend on the albumin concentrations.
The Staverman reflection coefficients for glucose
 in tissue and in the capillary wall
 are equal to  $\sigma_{TG}=\sigma_{G}=0.001$.
 Hereafter, the  values of other
parameters and absolute constants are taken from   Table 1.

Fig. 1 presents the spatial distributions of the steady-state density of fluid flux
from blood to tissue $q_U$ and the fluid flux across  tissue $j_U$, calculated
using  formulae (\ref{2.17})--(\ref{2.20}). The negative sign of $j_U$ indicates the net fluid flux occurs  across the tissue towards the peritoneal cavity. Therefore it corresponds to the water removal by ultrafiltration.
The monotonically decreasing (with the distance from the peritoneal surface)  function $q_U(x)$ and the  monotonically increasing
  function $j_U(x)$
 are in agreement with the
experimental data and previously obtained numerical results  for the
models that took into account only the glucose transport
(Cherniha et al. 2007; Waniewski et al. 2007).
 It should be stressed that in those previous models albumin transport was not considered and restrictions (\ref{2.13}) and
(\ref{2.16}) were not used.

Using the value of  the fluid flux $j_U$ at the point $x=0$,
one may calculate the reverse water flow (i.e. out of the tissue to
the cavity). Total fluid outflow from the tissue to the cavity (ultrafiltration),
calculated assuming that the surface area of the contact between
dialysis fluid and peritoneum is equal to $5\cdot10^3$ $cm^{2}$ (this surface area measured in 10 patients on peritoneal dialysis was found to be within the range from $0.41$ to $0.76$ $ m^2$ (Chagnac et al. 2002)), is about
$0.90$ $mL/min$. Note a similar value was   obtained previously in
(Cherniha et al. 2007) using numerical simulations. Moreover, it comes
 from formula  (\ref{1.3}) for  $x=0$  that  the ultrafiltration
increases with growing  $\sigma_{TG}$.
For example, if one sets $\sigma_{TG}=0.01$  into Eq.  (\ref{1.3}) then the  total fluid outflow from the tissue to the cavity is $5.2$ $mL/min$, what is very close to the value obtained in  (Cherniha et al. 2007) for the same parameters.

Figure 2 presents the spatial distributions of the glucose
concentration in   tissue
for $\sigma_{TG}=0.001$  and $\sigma_{TG}=0.01$ (see, Fig 2,  left picture). The interstitial glucose  concentration $C_G$ decreases rapidly
 with the distance from the peritoneal surface to the constant steady-state value of $C^*_G$  (see   formula (\ref{2.*}))  in
the deeper tissue layer  independently of the $\sigma_{TG}$ values  and is
practically $C^*_G$  for $x>0.3 $ (see the right picture, where both curves coincide). Thus, the width of the tissue layer with the increased glucose concentration (that is around 0.3 cm) does not depend on $\sigma_{TG}$.
This remains in agreement with the previous results
obtained in
 (Cherniha et al. 2007).

 We may conclude that, although the restriction  in the form of assumption
 (\ref{2.16})  is rather artificial from physiological point of view, the analytical formulae derived in Section 3 lead to the  results, which are similar to those obtained earlier with numerical simulations of pure glucose and water peritoneal transport (Cherniha et al. 2007), where this assumption  was not used.

Let us now consider  the second  case, which is  more realistic, i.e.,
 hereafter  restrictions (\ref{2.13}) and (\ref{2.25})
take place.

\medskip
\textbf{Remark.} In the case $\sigma_{TA}=0.0$ the results obtained via   formulae (\ref{2.29}) -- (\ref{2.32})  and ODEs  (\ref{2.33}) -- (\ref{2.34})
practically  coincide with those presented above (see Fig.1).
\medskip

Now we assume
that the Staverman reflection coefficient for albumin is different from zero  and equal to $\sigma_{A}=\sigma_{TA}=0.5$,  i.e.  the maximum value of $ \sigma_{TA}$ (see Table 1) is taken. In other words,  we assume that  the oncotic pressure plays an important role in contrary to the previous case.    In this case
the fluid flux across  tissue $j_U$  and the fluid flux
from blood to tissue $q_U$ (see formulae  (\ref{1.3}) and
(\ref{1.4}))  depend on  the interstitial concentrations of glucose and albumin.
 We performed many calculations using formulae (\ref{2.29}) -- (\ref{2.32})  and ODEs  (\ref{2.33}) -- (\ref{2.34})  for  a wide range of  values of the parameter  $\sigma_{TG}$,  including very small    ($0.001$) and   large  ($0.03$) those.
 Of course, some other parameters can vary as well, however, we restricted ourselves on this parameter because it is included in assumption   (\ref{2.13}).

The  results, obtained  for  $\sigma_{TG}=0.001$, $\sigma_{TG}=0.002$  and  $\sigma_{TG}=0.01$
are presented in Figs. 3 and 4. It is quite interesting that the
profiles for functions $q_U(x)$ and $j_U(x)$ shown in Fig. 3
are very similar to those in Fig. 1, although the relevant
formulae  are essentially different (the reader may compare
(\ref{2.29}) -- (\ref{2.32}) with (\ref{2.17})--(\ref{2.20})) and $\sigma_{TA}=0.5$.
Moreover, the form of these profiles are the same for a wide range  of the values of $\sigma_{TG} $.

\newpage

\textbf{Table 1.}
Parameters of the model used for numerical analysis of peritoneal transport.
\begin{small}
\begin{center}
\begin{tabular}{ll}
\hline


 Parameter name & Parameter symbol, value and unit \\


\hline &\\
 Minimal  fractional void volume  &
$\nu_{\min}=0.17 $
\\
 Maximal  fractional void volume  &
$\nu_{max}=0.35$   \\
Staverman reflection coefficient for glucose
  &
$\sigma_{TG}$   varies from $0$ to $0.03$ \\
Sieving coefficient of glucose in tissue &
 $S_{TG} = 1-\sigma_{TG}$  \\
Staverman reflection coefficient for albumin
&
 $\sigma_{TA}$  varies from
$0.05$ to $0.5$ \\
Sieving coefficient of
albumin in tissue &
 $S_{TA} = 1-\sigma_{TA}$ \\
 Hydraulic permeability of tissue &$K=5.14 \cdot 10^{-5}$ $cm^{2} \cdot min^{-1} \cdot mmHg^{-1}$\\
 Gas constant times temperature & $RT=18 \cdot 10^{3}$ $mmHg \cdot
mmol^{-1} \cdot mL$\\
 Width of tissue layer &  $L=1.0$ $cm$\\
 Hydraulic permeability of
capillary wall & $L_{P}a=$\\
times  density of capillary surface area & $7.3 \cdot 10^{-5}$
 $min^{-1} \cdot mmHg^{-1} $ \\
 Volumetric fluid flux  to
lymphatic vessels & $q_l =0.26\cdot 10^{-4}$  $min^{-1}$
 \\
 Diffusivity of glucose in tissue divided by $\nu_{min}$ &
 $D_G=12.11 \cdot 10^{-5}$ $cm^{2} \cdot min^{-1}$\\
 Diffusivity of albumin  in tissue divided by $\nu_{min}$  &
 $D_A=0.2
\cdot 10^{-5}$ $cm^{2} \cdot min^{-1}$ \\
 Permeability of capillary
wall for glucose &
$p_{G}a= $   \\
times  density of capillary surface area &$3.4 \cdot 10^{-2}$
$min^{-1}$ \\
Permeability of capillary wall for albumin &
 $p_{A}a=$
 \\
times  density of capillary surface area & $6\cdot 10^{-5}$
$min^{-1}$\\
 Glucose
concentration in blood &
 $C_{GB}=6\cdot 10^{-3}$ $mmol \cdot mL^{-1}$ \\
  Albumin
concentration in blood &
$C_{AB}=0.6\cdot 10^{-3}$ $mmol \cdot mL^{-1}$ \\
 Glucose
concentration in dialysate &
 $C_{GD}=180\cdot 10^{-3} $ $mmol\cdot mL^{-1}$ \\
 Albumin
concentration in dialysate &
  $C_{AD}=0$  \\
  Hydrostatic pressure of blood &
 $P_{B}=15$ $mmHg$  \\
 Hydrostatic pressure of dialysate &
 $P_{D}=12$ $mmHg$ \\
  Non-dimensional parameter &
$ \alpha=0.8 $ \\
\\ \hline
\end{tabular}
\end{center}
\end{small}

\begin{figure}[t]
\begin{minipage}[t]{5.5cm}
\centerline{\includegraphics[width=6.0cm]{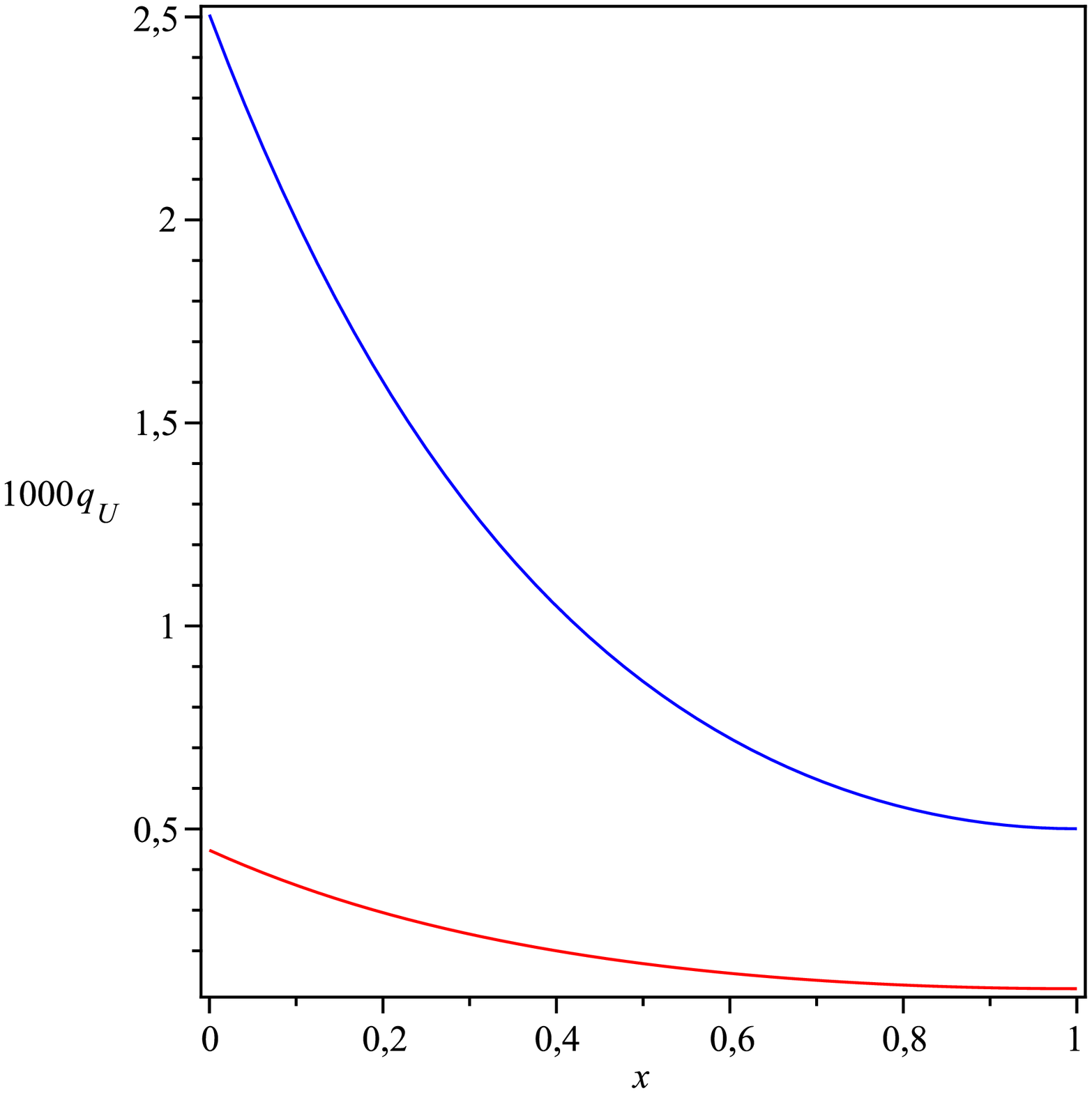}}
\end{minipage}
\hfill
\begin{minipage}[t]{5.5cm}
\centerline{\includegraphics[width=6.0cm]{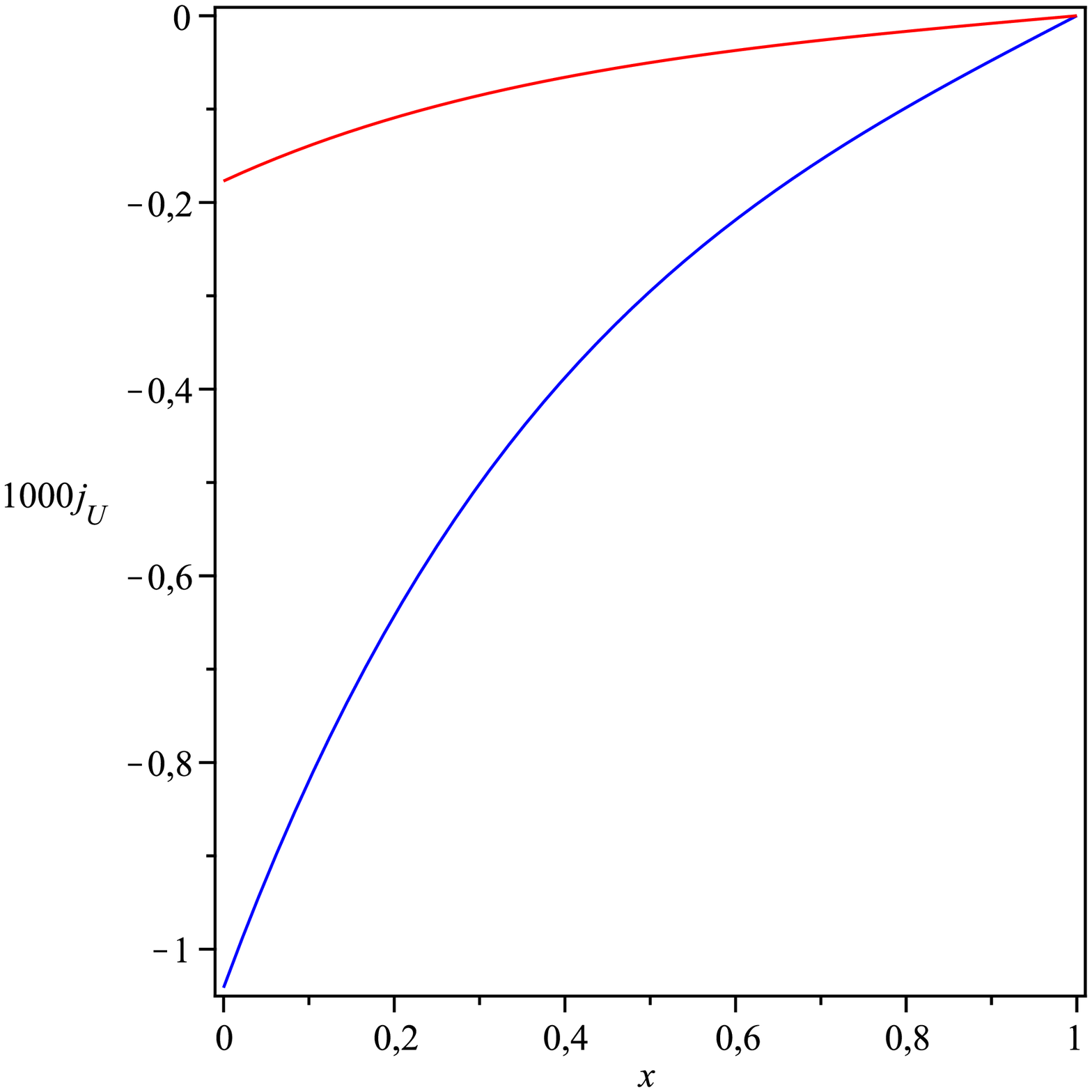}}
\end{minipage}
\caption{The
 fluid fluxes from blood to tissue $q_{U}$ (in $ min^{-1}$) and    across  tissue
 $j_{U}$ (in $ min^{-1}
\cdot cm$ ) as a function of distance from peritoneal cavity $ x$ (in $cm$)  for  $\nu= (\nu_{max}+\nu_{min})/2 $,
$\sigma_{TG}=0.001$ (red curve),   $\sigma_{TG}=0.01$ (blue curve),  and $\sigma_{TA}=0.0$.}
\label{Fig-1}
 \end{figure}

\begin{figure}[t]
\begin{minipage}[t]{5.5cm}
\centerline{\includegraphics[width=6.0cm]{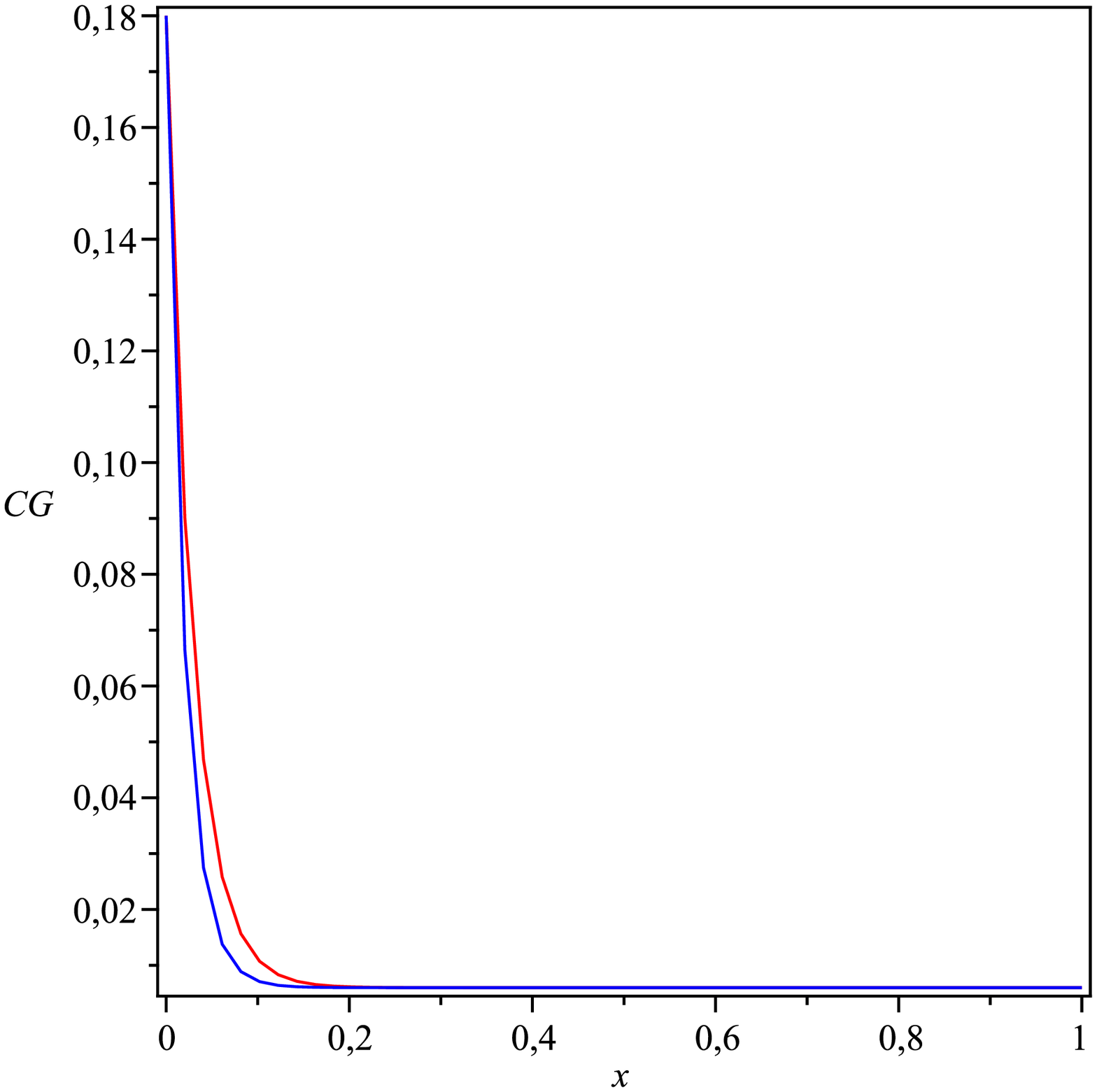}}
\end{minipage}
\hfill
\begin{minipage}[t]{5.5cm}
\centerline{\includegraphics[width=6.0cm]{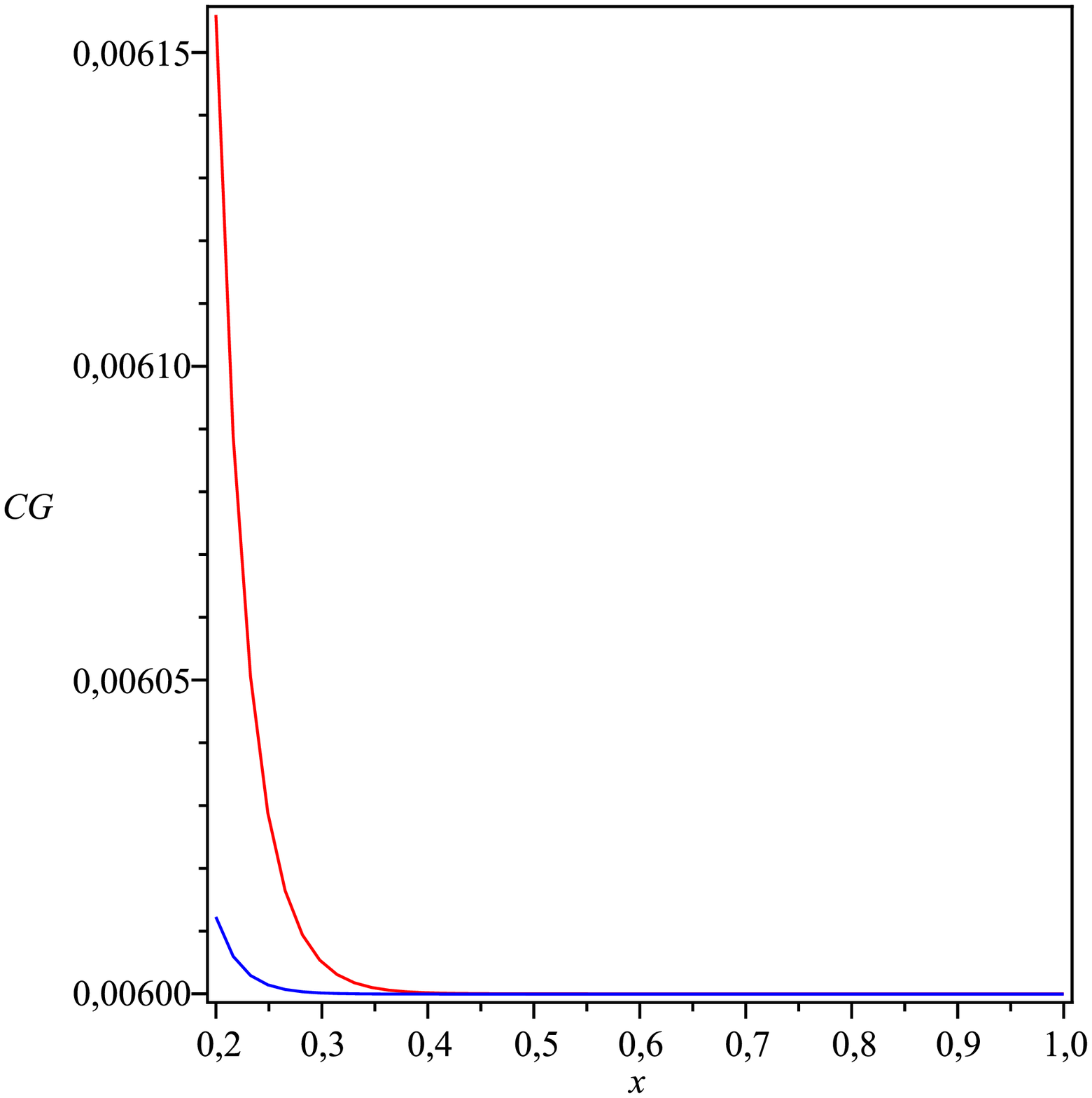}}
\end{minipage}
\caption{The
profiles of  glucose concentration in the tissue $C_G$ (in $mmol \cdot mL^{-1}$) as a function of distance from the peritoneal cavity $ x$ (in $cm$)  for $\nu= (\nu_{max}+\nu_{min})/2$,
$\sigma_{TG}=0.001$ (red curve)  and   $\sigma_{TG}=0.01$ (blue curve). Left panel for $0 \leq x  \leq1$, right panel: zoom for $0.2 \leq x \leq 1$.}
\label{Fig-2}
 \end{figure}


 \begin{figure}[t]
\begin{minipage}[t]{5.5cm}
\centerline{\includegraphics[width=6.0cm]{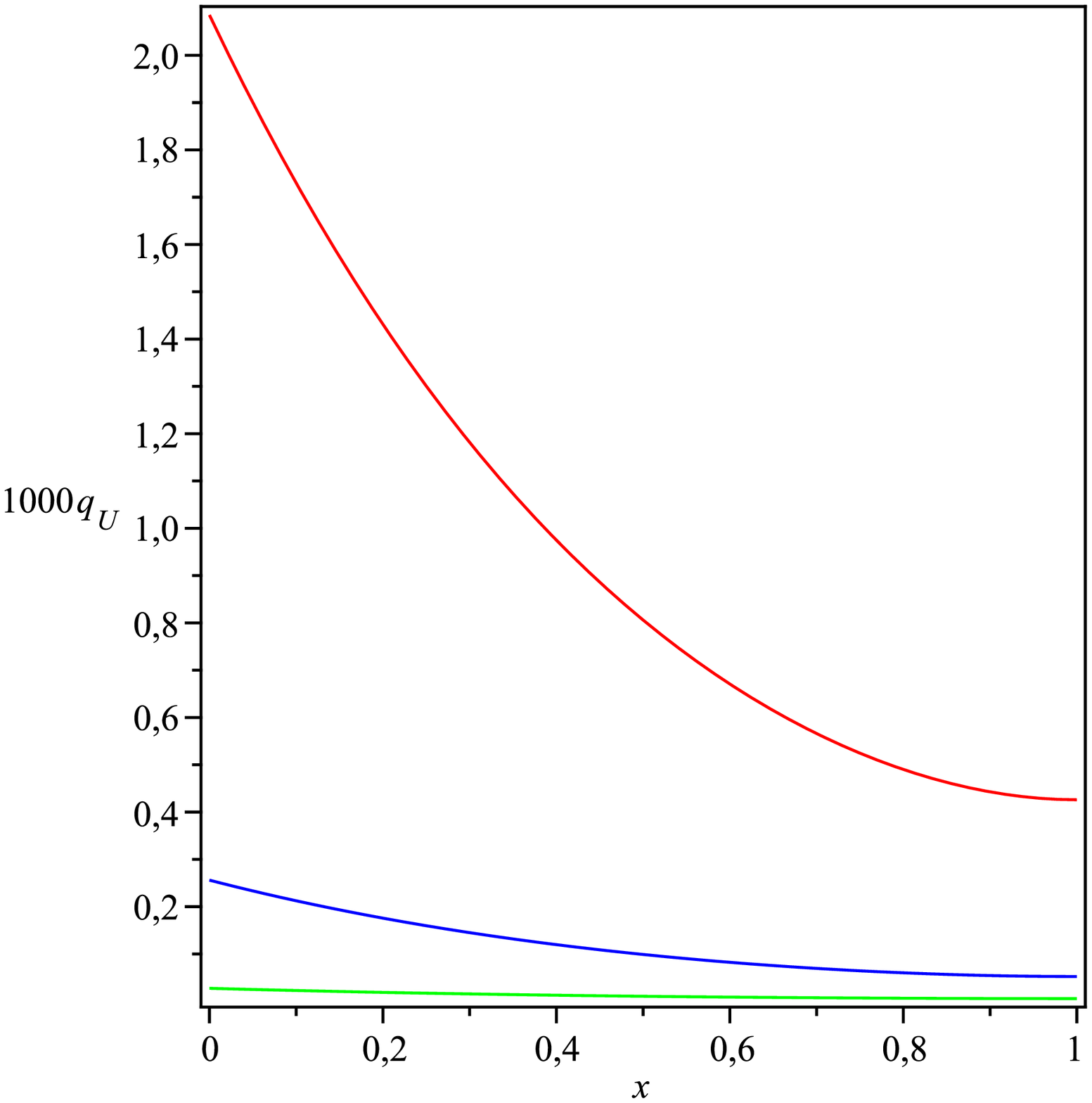}}
\end{minipage}
\hfill
\begin{minipage}[t]{5.5cm}
\centerline{\includegraphics[width=6.0cm]{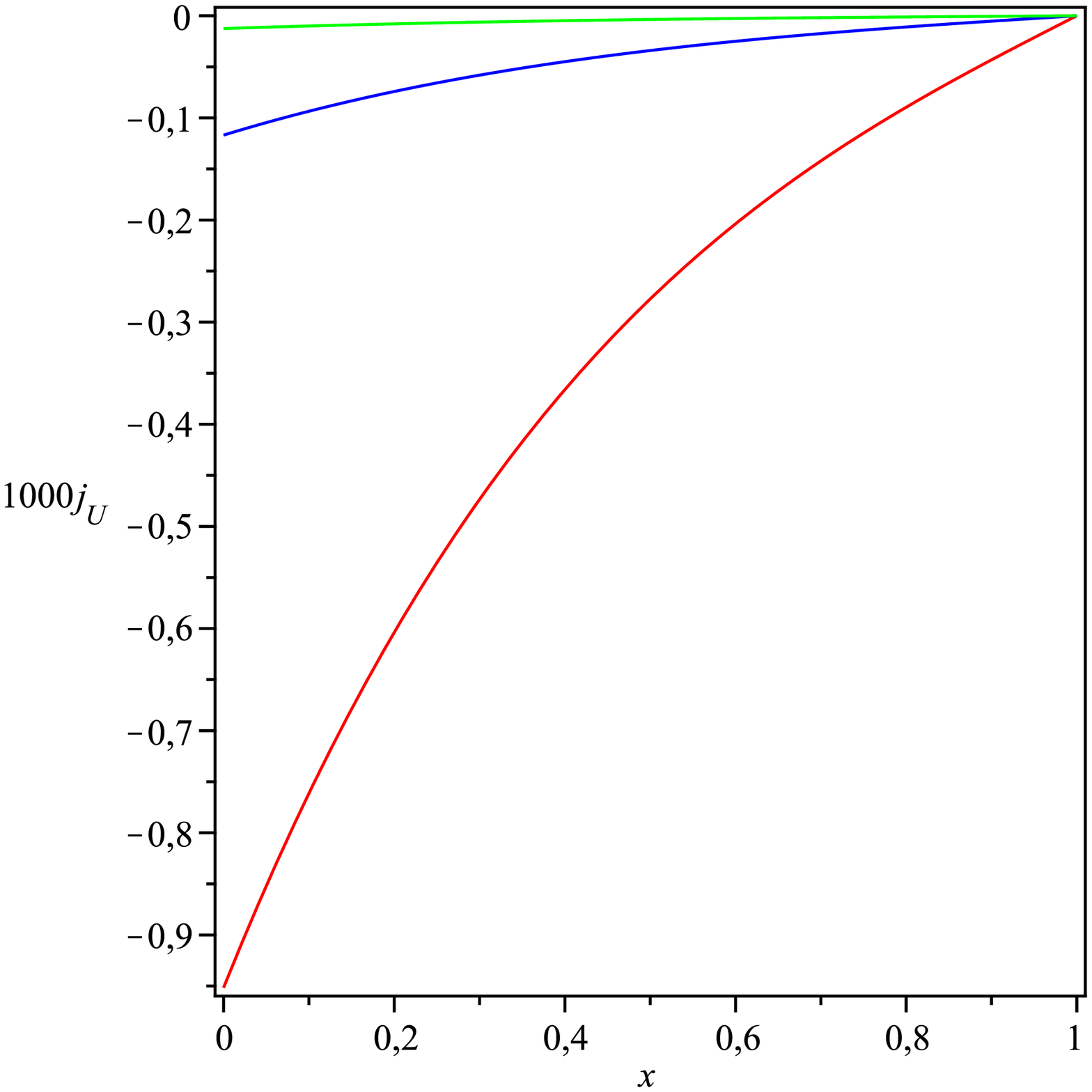}}
\end{minipage}
\caption{The
  fluid flux from blood to tissue $q_{U}$ (in $ min^{-1}$) and  the  fluid flux  across  tissue
 $j_{U}$ (in $ min^{-1}
\cdot cm$ ) as a function of distance from the peritoneal cavity $ x$ (in $cm$) for
 $\nu=\nu_{max} -(\nu_{max}-\nu_{min})x$,
  $\sigma_{TA}=0.5$, and $\sigma_{TG}=0.001$ (green); $0.002$ (blue); $0.01$ (red). }
\label{Fig-3}
 \end{figure}

\begin{figure}[t]
\begin{minipage}[t]{5.5cm}
\centerline{\includegraphics[width=6.0cm]{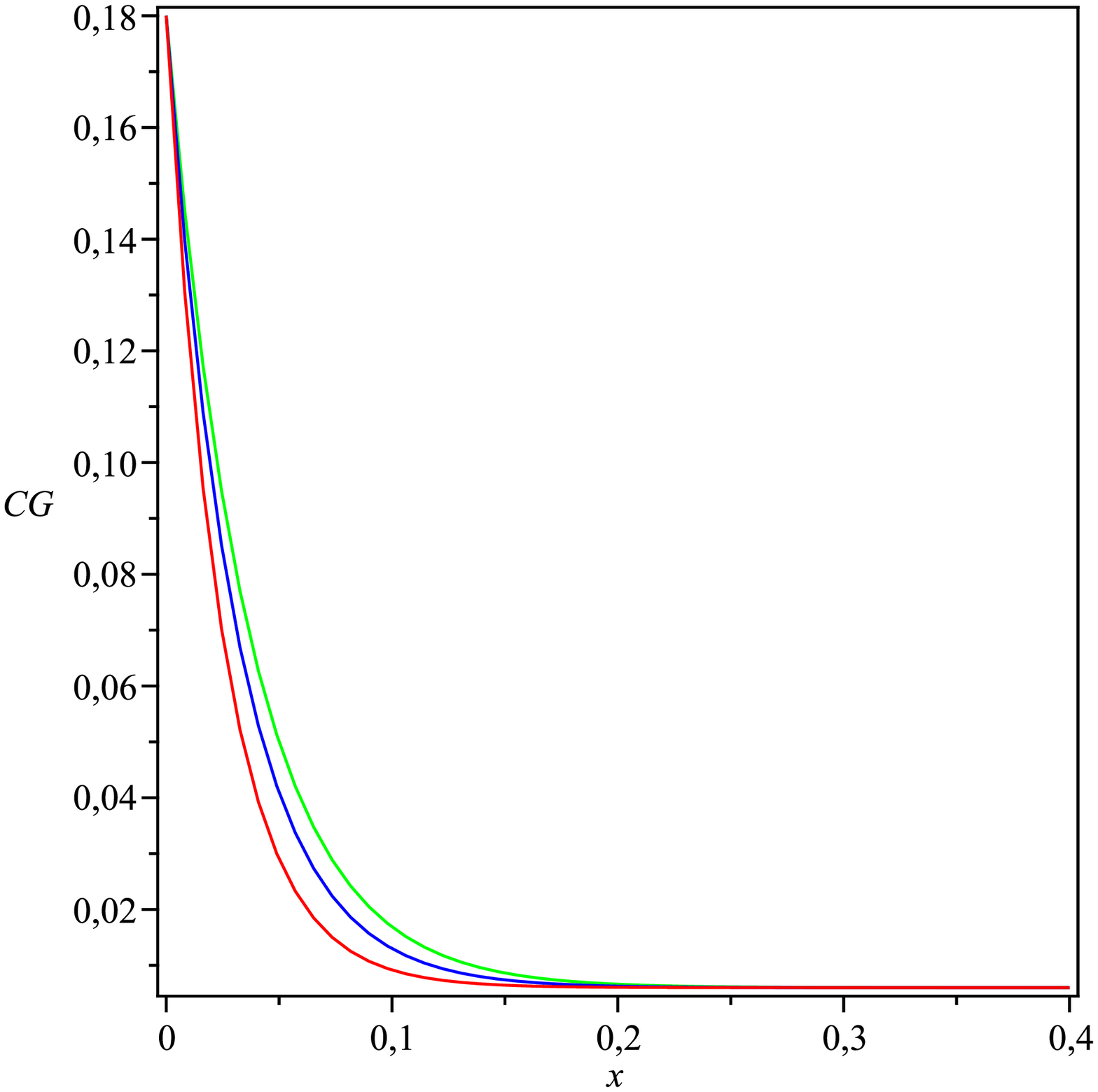}}
\end{minipage}
\hfill
\begin{minipage}[t]{5.5cm}
\centerline{\includegraphics[width=6.0cm]{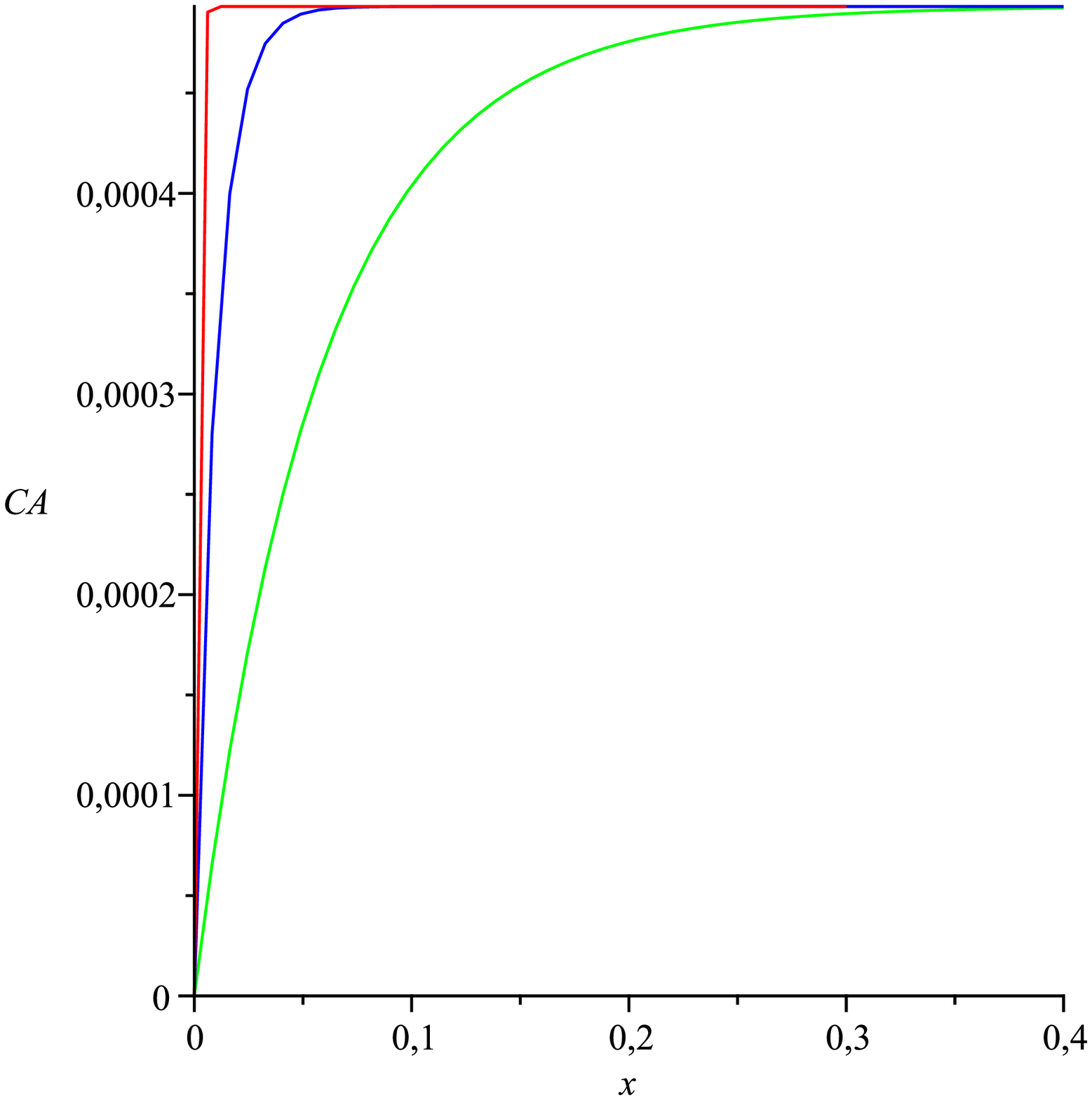}}
\end{minipage}
\caption{The
glucose concentration profiles, $C_G$ (in $mmol \cdot mL^{-1}$), and albumin concentration profiles, $C_A$ (in $mmol \cdot mL^{-1}$) , in the tissue as a function of distance from the peritoneal cavity, $ x$ (in $cm$), for  $\nu=\nu_{max} -(\nu_{max}-\nu_{min})x$,
 $\sigma_{TA}=0.5$, and $\sigma_{TG}=0.001$ (green); $0.002$ (blue); $0.01$ (red).   }
\label{Fig-4}
 \end{figure}


Using the value of  the fluid flux $j_U$ at the point $x=0$,
one may again  calculate the  ultrafiltration flow to the peritoneal cavity that can be obtained under the assumed here restrictions on the model parameters.
 In the  case $\sigma_{TG} =0.01$ the total fluid outflow from the tissue to the cavity  is approximately
equal to $4.8$ $mL/min$, while  it is very small ($0.06$ $mL/min$) for $\sigma_{TG} =0.001$.
To obtain   the values of the  ultrafiltration corresponding to those
 measured  during  peritoneal dialysis, we need to set    $\sigma_{TG} \geq0.02$.
   For example, setting $\sigma_{TG} =0.02$ and $\sigma_{TG} =0.03,$
 we obtain the   ultrafiltration  $10 $ $mL/min$  and  $15 $ $mL/min$, that are close to those measured in clinical conditions for similar boundary concentration of glucose   (Smit et al 2004a, Smit et al 2004b,
 Waniewski et al 1996a, Waniewski et al  2009,
  Stachowska et al  2012).  The initial values of ultrafiltration remain in agreement with values obtained by our group in clinical studies (Waniewski et al 1996b; Waniewski 2007). The initial rates of ultrafiltration for  3.86\% glucose solution were found to be of
$15 $ $mL/min$, which was much higher than those for 2.27 and 1.36\% glucose solutions, 8 and 6 ml/min, respectively (Waniewski et al 1996a,Waniewski et al 1996b). Similar values in the range of $14-18 $ $mL/min$ were measured during the initial minute and much lower values of $4-8 $ $mL/min$ at the end of a 4-h dwell study with 3.86\% glucose solution (Smit et al. 2004a; Smit et al. 2004b).

The spatial distributions of the glucose and albumin concentrations for different  values of  $\sigma_{TG}$
are pictured in Fig. 4. Note   that  glucose  concentration in the tissue, $C_G$,  again  decreases rapidly
 with the distance from the peritoneal cavity to the constant steady-state $C^*_G$    in
the deeper tissue layer.   The glucose concentration  is practically equal  to  $C^*_G$  for  any $x>0.1 $  $cm$ if   $\sigma_{TG}$ is large ($\geq 0.01$).
 The tissue layer with non-constant  $C_G$
is  slightly  wider  if  $\sigma_{TG}$  is small ($\leq 0.002$).   For such  values of  $\sigma_{TG}$, the glucose concentration  is  equal  to  $C^*_G$  for  any $x>0.2$ $ cm. $

The albumin
concentration in the tissue, $C_A$, is decreased ( in the direction to the peritoneal cavity) in a thin layer,  whereas it remains unperturbed  in the deeper tissue layers (Fig. 4). This decrease corresponds  to the transport of albumin to the peritoneal cavity that is most pronounced close to the peritoneal surface.
 We found that the albumin concentration   essentially
depends on the parameter $\sigma_{TG}$. In fact,  three curves presented
in Fig. 4
show that  the tissue layer with decreased   $C_A$ is wide  for $\sigma_{TG}=0.001$, whereas it much smaller for $\sigma_{TG}=0.002$
and almost vanishes for  $\sigma_{TG}=0.01$.
In the case $\sigma_{TG}=0.001$, the tissue layer of decreased
  $C_A$  is
about $0.3$ $cm$ indicating the removal of  albumin
 from this part of the tissue.
In the case of high $\sigma_{TG}$,
 the albumin concentration in the tissue
 is decreased only   in a very
thin layer, while  it remains  at physiological level  and  equal to $C^*_A$  (see   formula (\ref{2.*}))
behind this layer. Thus, high ultrafiltration flow contributes to the fast inflow of albumin from blood to the tissue and drags albumin from deep to subsurface layers. However, the diffusive leak of albumin from the tissue to the peritoneal cavity is faster with high ultrafiltration because of higher concentration gradient (Fig. 4).


\centerline{\textbf{5. Conclusions}}
\medskip
In this paper, a new  mathematical model for fluid transport  in
peritoneal dialysis was  constructed. The model is based on a
three-component nonlinear system of two-dimensional partial
differential equations and the relevant boundary and initial
conditions. To analyze the non-uniform  steady-state solutions, the
model was reduced to the non-dimensional form.
Under additional assumptions
the problem
was simplified in order to  obtain
analytical solutions in an explicit form.
As the result, the exact formulae for the  density of
fluid flux from blood to tissue  and the fluid flux across the
tissue were constructed together with two linear autonomous  ODEs for
glucose and albumin concentrations in the tissue.

The analytical results were checked for their  applicability to
describe the fluid-glucose-albumin  transport  in peritoneal dialysis.
 The selected values of the parameters were based on previous experimental and clinical studies or estimated from the data using the distributed model. Some of the parameters (Staverman reflection coefficients) were varied to check their impact on the model predictions. The model presented in the current study was extended, compared to the previous studies, by including the transport of water and two most important solutes related to water transport: glucose that is used as osmotic agent and albumin that is the primary determinant of oncotic pressure. These two solutes differ much (300 times) in molecular mass and therefore also differ in their transport parameters. The other studies include mostly only one of these two solutes into the model (Flessner et al. 1984; Baxter and Jain 1989,1990,1991;  Cherniha and Waniewski 2005; Flessner 2006; Cherniha et al. 2007; Waniewski et al. 2007,2009; Stachowska-Pietka et al. 2007,2012). On the other hand, our investigations are restricted to the steady state solutions, whereas in real dialysis the fluid and solute transport changes because of the change in boundary conditions (Stachowska-Pietka et al. 2006). We did not included into the model the phenomena of vasodilation and change in tissue hydration that yield spatially non-uniform structure of the tissue (however, our x-dependent fractional volume of interstitial fluid $\nu$ takes into account a part of this non-uniformity of the structure)  and contribute to the details of numerical solutions as compared to clinical data (Smit et al 2004a).
Only some of the model predictions can be compared directly to clinical data. The most important for us is the rate of ultrafiltration of water to the peritoneal cavity that is induced by glucose. With the concentration of glucose applied in our calculations the ultrafiltration rate of about 15 $mL/min$ is expected (Heimb\"urger et al. 1992; Waniewski et al. 1996a,1996b; Smit et al 2004a; Smit et al 2004b). Our results demonstrate that this value can be obtained if reflection coefficient for glucose is high (0.02 - 0.03). This general observation is in agreement with the early measurements of these coefficients, but differs from much higher values of the coefficient estimated previously by numerical simulations (Stachowska-Pietka et al. 2006; Waniewski  2007; Waniewski et al. 2007). The difference might be explained by the difference between glucose reflection coefficients for the tissue (low) and the capillary wall (high) obtained from previous numerical simulations, whereas these two coefficients were equal (with a medium value to yield the demanded ultrafiltration rate, for the sake of mathematical tractability) in our predictions (see below).

The glucose and albumin profiles obtained from our model are similar to those found in experimental studies (no such data are available for humans) and to the previous numerical simulations of clinical dialysis (Flessner et al. 2006;Waniewski et al. 2009; Stachowska-Pietka et al. 2012). The glucose interstitial concentration sharply decreases within 2 $mm$ from the peritoneal surface and is equal to its blood concentration in deeper tissue layers, see Figures 2 and 4, as it found previously in other numerical studies (Waniewski et al. 2009; Stachowska-Pietka et al. 2012). A similar profile was found in experiments performed in rats with manitol, which has identical transport characteristics as glucose (Flessner et al. 2006). In contrast, the interstitial concentration of albumin is high within the deep layers of tissue and decreases sharply in a thin layer close to the peritoneal tissue, Figures 2 and 4. This low subperitoneal protein concentration (and therefore also oncotic pressure) was confirmed experimentally (Rosengren et al. 2004) and in numerical simulations (Stachowska-Pietka et al. 2007). Some other results, as the profiles of
the flux from blood to tissue
 $q_U$, and the  flux  across tissue, $j_U$ ( see Figures 1 and 3), do not have any experimental counterpart and are rarely presented as the results of numerical studies. 

Thus, our model, although aimed at the investigation of its mathematical structure with specific coefficient conditions, yielded also some interesting predictions, in spite of rather simple approximations for the fractional interstitial fluid volume $\nu$ that were applied. Even the simplest approximations of $\nu$ by a constant or a linear function yielded the predictions in agreement with the models based on nonlinear dependence of $\nu$ on interstitial hydrostatic pressure.
In fact,  the monotonically decreasing (with the distance from the peritoneal surface)  function $q_U(x)$, describing  fluid flux
from blood to tissue,  and the  monotonically increasing
  function $j_U(x)$, describing fluid flux
across  tissue,
 are in agreement with the
experimental data and previously obtained numerical results.   Moreover, we calculated  the fluid flux $j_U(t,x)$
 at $ x = 0$, which   describes the net ultrafiltration flow, i.e.,  the efficiency of removal of water   during peritoneal dialysis, because it is important from practical point of view.
 The results show that the Staverman reflection coefficient for glucose $\sigma_{TG} $ plays the crucial role for the ultrafiltration.
 To obtain   the values of the  ultrafiltration corresponding to
experimental data, $10 - 20 $ $mL/min$,
 measured  during  peritoneal dialysis, we need to set    $\sigma_{TG} \geq0.02$ in the formulae  obtained.

The finding that high ultrafiltration flow rates measured in clinical studies may be obtained with relatively low $\sigma_G$ of  $0.01 - 0.03$ and at the same time rather high $\sigma_{TG} = \sigma_G$ (which is the assumption necessary to get the presented above analytical solutions) is interesting.   In fact,   much higher values of  $\sigma_{G}$ (about 0.5) and lower values of  $\sigma_{TG}$ (about 0.005)  were used in  (Waniewski et al. 2009, Stachowska-Pietka et al.  2012) to obtain similar  flow rates.
  The new solutions constructed  above
 add new perspective to the unsolved problem of the values of $\sigma_{G}$
   (see the detailed  discussion  in (Waniewski et al. 2009, Stachowska-Pietka et al.  2012)). Thus, these  new results are worth to be pursuit further not only because of mathematical interest but also of their potential practical applications.

 The difference between the present analytical solutions and the previous simulations is also in the profile fluid void volume being the outcome of the simulations whereas here this profile (approximated due to the linear function)  is an input to the equations.
 Other  approximations of  the fractional fluid volume $\nu$ may in future result in similar exact formulae. In the particular case, the preliminary calculations show that   such exact formulae can be obtained  when   $\nu$ is a decreasing exponential function.
 However, the assumption about the equality of the reflection coefficients in the tissue and in the capillary wall, which demonstrates an interesting specific symmetry in the equations, can be too restrictive for practical applications of the derived formulae
  (Waniewski et al. 2009). Therefore, other approaches to find  the analytical solutions of the model need to be looked for.


 \centerline{\textbf{  Acknowledgments.}}
\medskip
This work was done within
 the joint
project "Mathematical modeling transport processes in tissue during peritoneal dialysis" between PAS and NAS of Ukraine.

\renewcommand{\refname}{References}

\end{document}